\documentclass[12pt]{article}
\usepackage{fullpage,epsfig,graphics,amsbsy,amssymb,cancel,slashed,mathrsfs}
\usepackage{psfrag,hyperref}
\usepackage{graphicx,color}
\usepackage{wrapfig}
\usepackage{subfigure}

\newcommand{\be}{\begin{eqnarray}}
\newcommand{\ee}{\end{eqnarray}}
\usepackage{amsmath}
\numberwithin{equation}{section}
\usepackage{caption}
\usepackage[nosort]{cite}
 \usepackage[bulletsep]{collref}

\def\N{{\cal N}}

\def\al{\alpha}
\def\eps{\epsilon}

\newcommand{\bea}{\begin{eqnarray}}
\newcommand{\eea}{\end{eqnarray}}  
\newcommand{\nn}{\nonumber}
\newcommand{\Tr}{\textrm{Tr}}

\newcommand{\NN}{\mathcal{N}}

 \newcommand{\sfrac}[2]{\mbox{$\frac{#1}{#2}$}}

\newcommand{\LL}{{\mathcal L}}

\newcommand{\sbkt}[1]{\left[#1\right]}
\newcommand{\bkt}[1]{\left(#1\right)}

\newcommand{\Zv}{Z_{{1-\text{loop}}}^{\text{vec}}}
\newcommand{\Zh}{Z_{{1-\text{loop}}}^{\text{hyp}}}
\newcommand{\tv}{\langle\beta,\sigma\rangle^2}

\begin{document}

\thispagestyle{empty}
\begin{flushright} \small
UUITP-08/17\\
MIT-CTP/4889
 \end{flushright}
\smallskip
\begin{center} \LARGE
{\bf One-loop tests of supersymmetric gauge theories on spheres}
 \\[12mm] \normalsize
{\bf  Joseph A. Minahan\  ${}^a$ and Usman Naseer\ ${}^b$} \\[8mm]
 {\small\it
 ${}^a$\  Department of Physics and Astronomy,
     Uppsala University,\\
     Box 516,
     SE-751 20 Uppsala,
     Sweden
      \\  
      
         ${}^b$\ Center for Theoretical Physics,\\
     Massachusetts Institute of Technology,\\
     Cambridge, MA 02139, USA.
   }
  
  \medskip 
   \texttt{ joseph.minahan@physics.uu.se}\\
   \texttt{unaseer@mit.edu}

\end{center}
\vspace{7mm}
\begin{abstract}
 \noindent   
 
 \end{abstract}
 
 We show that a recently  conjectured form for perturbative supersymmetric  partition functions on  spheres of general dimension $d$ is consistent with the flat space limit of 6-dimensional $\NN=1$ super Yang-Mills.  We also show that the partition functions for $\NN=1$ 8- and 9-dimensional theories are consistent with their known flat space limits.

\eject
\normalsize

\tableofcontents

 \section{Introduction}

It has been long  known that there are difficulties in putting certain supersymmetric theories on Euclidean spaces  \cite{Zumino:1977yh}.  For instance, a single vector multiplet in a 4d $\NN=1$ gauge theory has a Majorana fermion as a superpartner to the gauge field in Minkowski space.  In rotating to Euclidean space it is impossible to maintain the reality condition on the fermion, so the supercharges are naturally complexified.   As shown by Zumino   \cite{Zumino:1977yh}, requiring the spinors to be real leads to extra supersymmetry with additional fields in the supermultiplet, namely two extra fermion degrees of freedom as well as two scalars, where one scalar has the wrong sign kinetic term.   In hindsight we can easily understand these extra fields as arising from a dimensional reduction of a Minkowskian six-dimensional vector multiplet, where one of the reduced directions is the time direction \cite{Blau:1997pp,Belitsky:2000ws}.

A similar issue occurs when we consider a vector multiplet in six dimensions.  In Minkowski space an $\NN=1$ vector multiplet has a real Weyl spinor.  Analytically continuing to Euclidean space complexifies the spinor.  In order to have real spinors one must increase the amount of supersymmetry to $\NN=2$ with 16 supersymmetries.  Now there will be  four scalars where one of them has the wrong sign kinetic term.  Note that these issues do not mean that we cannot have minimal supersymmetry on Euclidean spaces, it just means we cannot have it with only real fields.

However, the standard localization procedure often starts with ten-dimensional super Yang-Mills and dimensionally reduces to super Yang-Mills on lower dimensional spheres \cite{Pestun:2007rz,Minahan:2015jta,Pestun:2016zxk,Pestun:2016qko}.  As such, the fields are real.  Nevertheless, a simple argument shows that it should be possible to put a theory with $\NN=1$ supersymmetry on $S^4$.  Suppose one has a $U(1)$ vector multiplet on $R^4$, with no chiral multiplets charged under the $U(1)$.  This theory is free, hence it is conformal.  Therefore one can make a conformal transformation to $S^4$ and preserve the supersymmetry, albeit with complex fields.   In any case, one can use the methods of Festuccia and Seiberg, starting with off-shell supergravity to put any $\NN=1$ theory on $AdS_4$ and analytically continue to $S^4$ \cite{Festuccia:2011ws}. 
However, the full partition function for a generic interacting $\mathcal{N}=1$ theory  has been shown to be  scheme dependent \cite{Gerchkovitz:2014gta}, questioning whether one could obtain any scheme independent results from a localized partition function.  Nonetheless, certain theories might have further symmetries, opening up the possibility of localizing the theory.  A prominent example of this type is the $\mathcal{N}=1^*$ theory where it was shown that one can  obtain unambiguous and  scheme independent quantities constructed from the partition function \cite{Bobev:2016nua}.
 However, there is no known way to localize an $\NN=1$ theory on $S^4$.  The difficulty arises in trying to construct a  positive definite localization term.  It is possible to construct  a positive definite $Q$-exact term from a supersymmetry generator, however the known generators that give such terms do not themselves close to a symmetry of the Lagrangian \cite{guido}\footnote{We thank Guido Festuccia for several discussions on this point.}. {This issue is also discussed in \cite{Terashima:2014yua}}.

For six-dimensional $\NN=1$ super Yang-Mills the situation is worse.  In the four dimensional minimal supersymmetric case one expects to be able to put the theory on $S^4$ because of the existence of an appropriate superalgebra, namely $OSp(1|4)$, which has four supercharges and a bosonic $SO(5)$ subalgebra corresponding to the isometry of $S^4$.  For six dimensions we would want a superalgebra with a bosonic $SO(7)$ subalgebra and 8 supercharges transforming in a spinor representation of $SO(7)$, but no such superalgebra exists.  The $F(4)$ supergroup has an 
$SO(7)\times SU(1,1)$
 bosonic subalgebra and 16 supercharges, hence this is appropriate for $\NN=2$ supersymmetry.

We encounter a similar problem when considering super Yang-Mills in eight and nine dimensions.  In this case there are sixteen supersymmetries, and again, one would like to be able to put these theories on spheres.  But once more there is no corresponding superalgebra with the appropriate bosonic subalgebras, namely $SO(9)$ and $SO(10)$ for $S^8$ and $S^9$ respectively\footnote{There are superalgebras with $SO(9)$ and $SO(10)$ bosonic subalgebras, for example,  $OSp(9|2)$ and $OSp(10|2)$.  However, these algebras have supercharges that transform in the vector representation of $SO(9)$ and $SO(10)$ respectively, which violates Nahm's spin-statistics criterion for the classification of allowable superalgebras \cite{Nahm:1977tg}.}. 
Even if one were able to define such theories it might be the case that their partition functions suffer from ambiguities just like $\mathcal{N}=1$ theories in four dimensions.
In the next section we will review explicitly the obstacles in putting these supersymmetric theories on the spheres.

Nevertheless, there are some indications that there are theories  akin to $\N=1$ six, eight and nine dimensional supersymmetric theories on spheres.   In particular, evidence was presented in \cite{Minahan:2015any} showing that the perturbative partition functions  for super Yang-Mills with 8 supersymmetries on $S^3$, $S^4$ and $S^5$ have a natural analytic continuation, such that one can continue up to six dimensions.  Likewise theories with 16 supersymmetries on $S^d$ with $d=3,4,5,6,7$ also have a natural analytic continuation which can then be continued up to $d=8,9$.  Although, we do not have an explicit construction of Lagrangians for these theories, it is reasonable to assume that in the decompactification limit, they reduce to usual gauge theories in flat space. The main objective of this paper is to demonstrate that the partition functions are consistent with this picture. These partition functions include a dependence on one-loop determinants. We show that in the decompactification limit these one-loop determinants  produce the well known physics of the flat space theories.

The paper is organized as follows: In section \ref{localizationrev}  we briefly review the localization procedure in \cite{Minahan:2015jta} and the analytic continuation in \cite{Minahan:2015any}. In section \ref{divergences} we compute one-loop divergences from the analytically continued expressions for one-loop determinants and compare them with well known results in the literature.
In section \ref{s-summary} we summarize our results and discuss some further issues. 

 \section{Review of localization and its analytic continuation}
 \label{localizationrev}
 In this section we review the localization procedure for gauge theories with eight supersymmetries on spheres.  We use the conventions in \cite{Minahan:2015jta}  which generalizes the procedure on $S^4$ \cite{Pestun:2007rz}, where one starts with $\NN=1$ super Yang-Mills in ten dimensions and dimensionally reduces, including along a time-like direction so that the unreduced dimensions are Euclidean.  We then review analytic continuation of the localized result to other dimensions.
 
 In ten dimensions the Lagrangian in flat space is given by  \cite{Brink:1976bc}
 \be\label{LL}
\LL= \frac{1}{g_{10}^2}\Tr\left(\sfrac12F_{MN}F^{MN}-\Psi\slashed{D}\Psi\right)\,,
 \ee
 where the indices run from $M,N=0,\dots 9$ and $\Psi^a$ is a Majorana-Weyl spinor in the adjoint representation.  The gamma matrices $\Gamma^{M}_{ab}$ and $\tilde\Gamma^{M\,ab}$ are real and symmetric.
The Lagrangian  is invariant under the
 supersymmetry transformations
 \be\label{susy}
 \delta_\eps A_M&=&\eps\,\Gamma_M\Psi\,,\nn\\
  \delta_\eps \Psi&=&\sfrac12 \Gamma^{MN}F_{MN}\,\eps\,,
 \ee
where $\eps$ is any constant real spinor.  We then dimensionally reduce to $d$ dimensions so that the gauge fields are $A_\mu$, $\mu=1,\dots d$, while the remaining bosonic fields are scalars with $\phi_I\equiv A_I$, $I=0,d+1,\dots 9$.  The scalar $\phi_0$ will have a wrong-sign kinetic term in the action because it came from the dimensional reduction of the time direction.  This also leads to a noncompact $R$-symmetry, $SO(1,9-d)$.  The dimensionally reduced field-strengths become
\be
 F_{\mu I}&=&[D_\mu,\phi_I]\nn\\
 F_{IJ}&=&[\phi_I,\phi_J]\,.
 \ee

Putting the $d$-dimensional Euclidean space on the sphere modifies the fermion supersymmetry transformation to
 \be\label{susysp}
  \delta_\eps \Psi&=&\sfrac12 \Gamma^{MN}F_{MN}\eps+\frac{\alpha_I}{2}\Gamma^{\mu I}\phi_I\nabla_\mu\,\eps\,,
 \ee
where the constants $\al_I$ are given by
\be\label{alrel}
\alpha_I&=&\frac{4(d-3)}{d}\,,\qquad I=8,9,0\nn\\
\alpha_I&=&\frac{4}{d}\,,\qquad I=d+1,\dots 7\, .
\ee
The index $I$ in (\ref{susysp}) is summed over.  The supersymmetry parameters $\eps$ are special cases of the superconformal Killing spinors on $S^d$ that satisfy
\be\label{KS1}
\nabla_\mu\eps=\tilde\Gamma_\mu\tilde\eps\,,\qquad\qquad \nabla_\mu\tilde\eps=-\beta^2\Gamma_\mu\eps\,.
\ee
where $\beta=\frac{1}{2r}$, which has 32 independent components.  We reduce this to 16 components by setting
\be\label{KS}
\nabla_\mu\eps=\beta\,\tilde\Gamma_\mu\Lambda\, \eps\,,
\ee
where $\Lambda$ must satisfy $\tilde\Gamma^\mu\Lambda=-\tilde\Lambda\Gamma^\mu$,  $\tilde\Lambda\Lambda=1$, $\Lambda^T=-\Lambda$ for consistency with (\ref{KS1}).  A minimal choice  is $\Lambda=\Gamma^0\tilde\Gamma^8\Gamma^9$, showing that this construction works up to $d=7$.  
If we write the sphere metric as 
\be\label{metric}
ds^2=\frac{1}{(1+\beta^2x^2)^2}\, dx_\mu dx^\mu\,,
\ee
then the solutions to (\ref{KS}) are given by
\be\label{spinsol}
\eps=\frac{1}{(1+\beta^2 x^2)^{1/2}}\left(1+\beta\, x\cdot\tilde\Gamma\,\Lambda\right)\eps_s\,,
\ee
where $\eps_s$ is a constant spinor.  The supersymmetric Lagrangian is given by
\be\label{Lss}
\LL_{ss}&=&\frac{1}{g_{YM}^2}\Tr\Bigg(\sfrac12F_{MN}F^{MN}-\Psi\slashed{D}\Psi+\frac{(d-4)}{2r}\Psi\Lambda\Psi+\frac{2(d-3)}{r^2} \phi^A\phi_A+\frac{(d-2)}{r^2}\phi^i\phi_i\nn\\
&&\qquad\qquad\qquad\qquad -\frac{2}{3r}(d-4)[\phi^A,\phi^B]\phi^C\varepsilon_{ABC}\Bigg)\,.
\ee
The scalars are split into two groups, $\phi^A$, $A=0,8,9$ and $\phi^i$, $i=d+1,\cdots 7$.  One can see from the structure of (\ref{Lss}) that the $R$-symmetry is generically broken from $SO(1,9-d)$ to
 $SO(1,7-d)$.

The Lagrangian in (\ref{Lss}) preserves 16 supersymmetries, but we can reduce this to $8$ for $d\le5$ by imposing the extra condition $\eps=+ \Gamma\eps$, where $\Gamma=\Gamma^{6789}$.  The fermions then split into the components of a vector multiplet $\psi$, with $\psi=+\Gamma\psi$, and a hypermultiplet $\chi$ with $\chi=-\Gamma\chi$.   The gauge fields $A_\mu$ and scalars $\phi_0$, $\phi_{d+1}\dots \phi_5$ are the bosonic fields of the vector multiplet, while $\phi_6,\dots \phi_9$ are the bosonic fields for the hypermultiplet.  With fewer supersymmetries we can add mass terms to the Lagrangian for the hypermultiplets.  This also modifies the $\alpha_I$ for the hypermultiplet scalars to 
\be\label{al8}
\al_I&=&\frac{2(d-2)}{d}+\frac{4i\sigma_I \,m\,r}{d}\qquad I=6\dots 9\nn\\
\sigma_I&=&+1\qquad\qquad I=6,7\nn\\
\sigma_I&=&-1\qquad\qquad I=8,9\,,
\ee
while at the same time modifying the extra terms in the Lagrangian to
\be\label{3phi}
\LL_{\phi\phi\phi}=\frac{1}{g_{YM}^2}\left(\left(\frac{2(d-4)}{r}+4im\right)\Tr(\phi^0[\phi^6,\phi^7])-\left(\frac{2(d-4)}{r}-4im\right)\Tr(\phi^0[\phi^8,\phi^9])\right)\,,
\nn\\
\ee
and the quadratic terms for the hypermultiplet to
\be\label{Lchi}
\LL_{\chi\chi}=\frac{1}{g_{YM}^2}\left(- im\Tr \chi\Lambda\chi\right)\,,\nn\\
\LL_{\phi\phi}= \frac{1}{g_{YM}^2}\left(\frac{d\,\Delta_I}{2\,r^2}\,\Tr \phi_I\phi^I\right)\,,
\ee
where
\be\label{Delta8}
\Delta_I=\frac{2}{d}\left(mr(mr+i\sigma_I)+\frac{d(d-2)}{4}\right)\,.
\ee

Partition functions for theories on $S^3,S^4,S^5,S^6,S^7$ were computed in 
\cite{Kapustin:2009kz,Pestun:2007rz,Kallen:2012va,Minahan:2015jta}. Partition functions for $S^d$,  can be written as
\be \label{eq:Zfull}
Z\ =\ \int \sbkt{d\sigma}_{\text{Cartan}} \exp\bkt{-\frac{8\pi^{\frac{d+1}{2}} r^{d-4} \sigma^2}{g^2\Gamma\bkt{\frac{d-3}{2}}}}\ \Zv \bkt{\sigma} \Zh\bkt{\sigma},
\ee
where $\Zv$ and $\Zh$ are vector- and hyper-multiplet one-loop determinants. It was observed in  \cite{Minahan:2015any} that the one-loop determinants of all known examples with eight supercharges on $S^d$, where $3\le d\le5$, can be written in the more general form
 \be 
\label{eq:Z1vec}
Z_{1-\text{loop}}^{\text{vec}}\ \bkt{\sigma}\prod_{\beta>0}\langle \beta,\sigma\rangle^2 =\ \prod_{\beta>0} \prod_{n=0}^{n=\infty}\ \sbkt{\bkt{n^2+\langle \beta,\sigma\rangle^2}\bkt{\bkt{n+d-2}^2+\langle \beta,\sigma\rangle ^2}}^{\frac{\Gamma\bkt{n+d-2}}{\Gamma[n+1]\Gamma[d-2]}}.
\ee
\be\label{eq:Z1hyp}
\Zh =\ \prod_{\xi} \prod_{n=0}^{n=\infty}\ \sbkt{\bkt{n+\frac{d-2}{2}}^2+\bkt{\langle \xi,\sigma\rangle+\mu} ^2}^{-\frac{\Gamma\bkt{n+d-2}}{\Gamma[n+1]\Gamma[d-2]}},
\ee
where 
$\xi$ are weights in the hypermultiplet representation and $\mu=m r$ is the dimensional-less mass parameter. Theories with sixteen supercharges are obtained by taking the vector multiplet with eight supercharges and adding a massless hypermultiplet in the adjoint gauge group. Hence the resulting one-loop determinants would be:
\be \label{eq:Zvh16}
\Zv\Zh \prod_{\beta>0} \tv\ =\ \prod_{\beta>0} \prod_{n=0}^{n=\infty}\ \sbkt{\frac{n^2+\tv}{\bkt{n+d-3}^2+\tv}}^{\frac{\Gamma\bkt{n+d-3}}{\Gamma\bkt{n+1}\Gamma\bkt{d-3}}}.
\ee
This, not only gives the correct expression for theories with sixteen supercharges in $3\leq d \leq 5$, but also agrees with the one-loop determinants on $S^6$ and $S^7$ for sixteen supercharges.

It now seems quite natural to analytically continue the dimension past $d=5$ in (\ref{eq:Z1vec}) and (\ref{eq:Z1hyp}) up to $d=6$, even though there is no appropriate $\Lambda$ and $\Gamma$ that can accommodate the above construction of vector and hypermultiplets on $S^6$.  At this point we will assume that our presumed theory has 8 supersymmetries and reduces to standard $\NN=1$ super Yang-Mills in the flat space limit.  One possibility is that the $SO(7)$ symmetry is broken to $SO(6)$ with the  spinors having opposite chiralities on the north and south poles.  

Likewise, it seems natural to analytically continue 
 the dimension past $d=7$ in (\ref{eq:Zvh16}), even though there is no appropriate $\Lambda$ for the construction.  In the $d=8$ and $d=9$ cases we will again assume that our presumed theory reduces to standard $\NN=1$ SYM in the flat space limit.  On the spheres we assume that the rotational symmetry is broken to $SO(8)$.  

  \section{One-loop divergences from partition functions}
  \label{divergences}
  In this section we will use the analytically continued expressions for one-loop determinants to compute effective couplings for theories with eight and sixteen supersymmetries in diverse dimensions. The ultraviolet divergences of the gauge coupling at one-loop can then be compared with the counter terms for supersymmetric theories at one-loop. In four dimensions, upon taking the decompactification limit one can compute the beta function of the theory. We show that results obtained from the analytically continued one-loop determinants are in agreement with explicit one-loop computations in these theories.
 \subsection{Eight supersymmetries in $4d$}
Recall that for a gauge theory in four dimensions with $N_f$ Dirac fermions in representation ${\bf R_{f}}$ and $N_s$ complex scalars in representation ${\bf R_s}$ of a semi-simple gauge group, the one loop beta function is given by
\be 
\beta\bkt{g}\ =\ -\frac{1}{16\pi^2} g^3\bkt{\frac{11}{3}C_2\bkt{\text{\bf Adj}}-\frac{4}{3} N_f C_2\bkt{\bf R_f}-\frac{1}{3} N_s C_2\bkt{\bf R_s}},
\ee
$C_{2}\bkt{\bf R}$ is the quadratic Casimir in the representation ${\bf R}$ of the gauge group. For $\NN=2$ theory with $N_h$ hypermultiplets in the representation ${\bf R}$ of the gauge group the beta function becomes
\be 
\label{eq:betaConv}
\begin{split}
 \beta\bkt{g}\ & =\ \frac{g^3}{8\pi^2} \bkt{N_h\ C_2\bkt{\bf R}-   C_2\bkt{\text{\bf Adj}}}.
\end{split}
\ee 
The contribution from the vector multiplet was previously found in \cite{Pestun:2007rz} by taking the  hypermultiplet mass to infinity in the $\NN=2^*$ theory.
We want to reproduce (\ref{eq:betaConv}) by using the analytically continued one-loop determinant for the vector and hyper multiplets given in equations (\ref{eq:Z1vec}) and (\ref{eq:Z1hyp}).  

To do so we need to determine $\mathcal{O}\bkt{\sigma^2}$ terms appearing in the one-loop determinants. To proceed we replace $\sigma$ by $t\sigma$ in the expressions for the one-loop determinants. The parameter $t$ keeps track of the order of $\sigma$. Focusing only on the vector multiplet,  one can easily find that
\be  
\begin{split}
\frac{d\log \Zv}{dt^2}\ + \sum_{\beta>0} \frac{1}{t^2}\ & =  \\ 
& \sum_{\beta>0}\tv\bkt{ \mathcal{F}\bkt{d-2,0,t\ \langle \beta,\sigma\rangle}\ + \mathcal{F}\bkt{d-2,d-2,t\ \langle \beta,\sigma\rangle}},
\end{split}
\ee
where
\be 
\mathcal{F}\bkt{x,y,z}\equiv\sum_{n=0}^{\infty} \frac{\Gamma\bkt{n+x}}{\Gamma\bkt{n+1}\Gamma\bkt{x}}\ \frac{1}{\bkt{n+y}^2 +z^2}\ =\ \frac{i}{2 z} \bkt{\frac{1}{y+i z} {}_2F_1 \bkt{x, y+i z; y+ iz+1;1}-c.c}.\nonumber
\ee
For $d=4-\epsilon$, we expand the R.H.S in powers of $t$ and $\epsilon$.  Keeping only the leading terms, we find
\be 
\frac{d\log \Zv}{dt^2}\ = \frac{2}{\epsilon} C_2\bkt{\text{\bf Adj}} \sigma^2+\cdots.
\ee
From this we can easily obtain
\be 
\log \Zv\ =\ \frac{2}{\epsilon} C_2\bkt{{\bf Adj }} \sigma^2\ + \cdots. 
\ee
A completely analogous calculation for a hypermultiplet in representation ${\bf R}$ of the gauge group gives
\be 
\log \Zh= -\frac{2}{\epsilon} \sigma^2 C_2\bkt{{\bf R}}+\cdots.
\ee
For a gauge multiplet and $N_h$ hypermultiplets, the contribution to the $\mathcal{O}\bkt{\sigma^2}$ term from the one-loop determinants can be combined with the $\mathcal{O}\bkt{\sigma^2}$ term in the fixed point action as given in equation (\ref{eq:Zfull}) to get
\be 
\frac{8\pi^2}{g^2\bkt{{\Lambda}}} = \bkt{ \frac{8\pi^2}{g_{0}^2}+ \frac{2}{\epsilon} C_2\bkt{{\bf Adj}}-\frac{2}{\epsilon} N_h C_2\bkt{\bf R} }{\Lambda}^{-\epsilon},
\ee
where $g\bkt{\Lambda}$ is the running coupling constant at   the renormalization scale $\Lambda\sim  r^{-1}$ \cite{Pestun:2007rz},  $g_0$ is the bare coupling. From the above equation one can easily obtain the beta function,
\be 
\begin{split}
 \beta\bkt{g}\ & =\ \frac{g^3}{8\pi^2} \bkt{N_h\ C_2\bkt{\bf R}-   C_2\bkt{\text{\bf Adj}}}.
\end{split}
\ee 
This matches precisely with equation (\ref{eq:betaConv}).
For one hypermultiplet in the adjoint representation the beta function vanishes. This is to be expected since it corresponds to $\mathcal{N}=4$ SYM. 

\subsection{Eight supersymmetries in 6d}
Since the explicit expression for one loop determinants for eight supersymmetries in $4d$ are known in terms of infinite products, the above results can be reproduced by regularizing those expressions by introducing a finite cut off parameter $\Lambda r$ and then taking the decompactification limit $r\to\infty$.  As explained earlier, it is not known how to localize a six dimensional theory with eight supersymmetries. In this case the expression (\ref{eq:Z1vec}) is a genuine ansatz. In this subsection we will perform a non trivial check on that ansatz by computing the effective coupling. It is well known that the six dimensional theory with eight supersymmetries has a quadratic divergence at one-loop \cite{Howe:1983jm,Marcus:1984ei}. We will compute the effective coupling using the one loop determinant (\ref{eq:Z1vec}) and show that it has a quadratic divergence in the decompactification limit.

Since dimensional regularization is only sensitive to logarithmic divergences we will use a hard cutoff to isolate the quadratic divergence. At leading order in the divergence this is expected to be consistent with supersymmetry.  However, there could be issues with sub-leading divergences, if for example imposing the cutoff leaves off the super-partners of modes at or near the cutoff.  However, assuming that the proposed dimensional regularization respects the supersymmetry we can show that the logarithmic divergences coming from the hard cutoff are consistent with the result coming from dimensional regularization, even if the log divergence is sub-leading.

We use $d=6$ in (\ref{eq:Z1vec}) and truncate the infinite product at $n_{{\bf max}}= \Lambda r$ to find quadratic dependence on the energy cutoff $\Lambda$. It is straightforward to find that the divergent contribution to the $\sigma^2$ term from the vector one-loop determinant is
\be 
\log \Zv\ =\ \bkt{\frac{\Lambda^2 r^2+ 5\Lambda r}{6}+ \frac{11}{3}\log\bkt{\Lambda r}} C_2\bkt{\bf Adj} \sigma^2\ + \cdots.
\ee

By combining this with the fixed point action, we find the effective coupling given by
\be 
\frac{16\pi^3 r^2}{g^2}\ =\ \frac{16\pi^3 r^2}{g_0^2} -\bkt{\frac{\Lambda^2 r^2+ 5\Lambda r}{6}+ \frac{11}{3}\log\bkt{\Lambda r}} C_2\bkt{\bf Adj}.
\ee
In the $r\to \infty$ limit only the leading terms in $r$ survive and one obtains
\be 
\frac{1}{g^2}\ =\ \frac{1}{g_0^2}-\frac{\Lambda^2}{96\pi^3} C_2\bkt{\bf Adj}.
\ee
We see that the effective coupling diverges quadratically with the scale $\Lambda$.

It is also known that the six dimensional theory can be made finite at one loop by adding a suitable hypermultiplet. This would be the case if the hypermultiplet and the vector multiplet contribute to the quadratic divergence with opposite sign. This is also consistent with the one-loop determinant (\ref{eq:Z1hyp}). For a hypermultiplet in representation $\bf R$, the contribution to $\mathcal{O}\bkt{\sigma^2}$ term is given by
\be 
\log \Zh\ =\  \bkt{-\frac{\Lambda^2 r^2+ 5\Lambda r}{6}+ \frac{1}{3}\log\bkt{\Lambda r}} C_2\bkt{\bf R} \sigma^2\ + \cdots.
\ee
So that the effective coupling with $N_h$ hypermultiplets in the representation $\bf R$ is given by:
\be 
\frac{1}{g^2}\ =\ \frac{1}{g_0^2}-\frac{\Lambda^2}{96\pi^3}\bkt{ C_2\bkt{\bf Adj}-N_h C_2\bkt{\bf R}}.
\ee
In particular, for a single hypermultiplet in the adjoint representation the quadratic divergence vanishes as expected \cite{Howe:1983jm}. 
\subsection{Sixteen supersymmetries in 4d and 6d}
In four and six dimensions, explicit expressions for the one-loop determinants for sixteen supersymmetries are known. We will compute the effective coupling at one-loop using both of these expressions and show that it is consistent. 

For four dimensions, we compute effective coupling from (\ref{eq:Zvh16}) using $d=4$. We will truncate the infinite product at $n_{\bf max}=\Lambda r$. By doing so one finds that the contribution to $\mathcal{O}\bkt{\sigma^2}$ term vanishes.
\be 
\log\bkt{\Zv\Zh}\ = 0+\cdots.
\ee
Hence the coupling is not affected by one-loop effects and is independent of the cutoff scale $\Lambda$.  Now we can easily check that this result is consistent with analytically continued expression, i.e., expanding (\ref{eq:Zvh16}) in powers of $\epsilon$ for $d=4-\epsilon$. Replacing  $\sigma $ by $t\sigma$ one can easily obtain
\be 
\frac{d}{dt^2} \log\bkt{\Zv\Zh}\ = \mathcal{O}\bkt{d-4} \implies 
\log\bkt{\Zv\Zh}\ =\ \mathcal{O}\bkt{d-4}.
\ee
Which is consistent with the result obtained from the explicit expression. 

Similarly, in six dimensions the contribution to the $\mathcal{O}\bkt{\sigma^2}$ term from (\ref{eq:Zvh16}) takes the  form
\be 
\log\bkt{\Zv\Zh}\ =\ 3\bkt{ \log \left(\Lambda r\right)-\frac{1}{2}+ \gamma} C_2\bkt{\bf Adj} \sigma^2+\cdots.
\ee
Combining this with the fixed point action, we obtain the effective coupling which is given by:
\be 
\frac{1}{g_{\bf }^2\bkt{\Lambda}}\ =\ \frac{1}{g_0^2}-\frac{3}{16\pi^3 r^2}\bkt{ \log \left(\Lambda r\right)-\frac{1}{2}+ \gamma} C_2\bkt{\bf Adj},
\ee
We see that the coupling has a logarithmic dependence on the scale $\Lambda$. It is easy to see that Logarithmic dependence is produced by using dimensional regularization for  $d=6-\epsilon$ in the analytically continued expression. Doing so we find that:
\be 
\log\bkt{\Zv\Zh}\ =\  \frac{3}{\epsilon} C_{2}\bkt{\bf Adj}\sigma^2+\cdots.
\ee
Combining this with the contribution from the fixed point action and noting that in $6-\epsilon$-dimensions $\frac{r^2}{g^2}$ has mass dimension $-\epsilon$ we get:
\be 
\frac{1}{g^2\bkt{\Lambda}}\ =\ \Lambda^{-\epsilon}\bkt{\frac{1}{g_0^2}+\frac{3}{16\pi^3 r^2 \epsilon}}\ =\ \frac{1}{g_0^2}-\frac{3}{16\pi^3 r^2} \log\bkt{\Lambda},
\ee
where a $\Lambda$-independent infinite piece is absorbed in $\frac{1}{g_0^2}$.
This gives the same logarithmic dependence on the energy scale $\Lambda$. Note that in the decompactification limit this logarithmic divergence vanishes, consistent with the fact that the six dimensional theory with 16 supersymmetries is finite at one-loop. 

\subsection{Sixteen supersymmetries in 8d and 9d}
For $d=8,9$, it is not known how to localize. Here we show that the analytically continued expression for the one-loop determinant is consistent with known results. It is known that for $d=8,9$, none of the terms present in the tree level Lagrangian need a counter term at one-loop \cite{Marcus:1984ei,Howe:2002ui}. Hence, the effective coupling determined from the analytically continued expressions for one-loop determinants should not have any divergences. This can be easily demonstrated by using the methods of this section. 
A short calculation shows that the contribution to $\mathcal{O}\bkt{\sigma^2}$ term from the one-loop determinant (\ref{eq:Zvh16}) for $d=8$ is
\be 
\log\bkt{\Zv\Zh}\ =\ \frac{5}{6}\bkt{\frac{ \bkt{\Lambda r}^2}{4}+\frac{3 \Lambda r}{2}+ 5 \log\bkt{\Lambda r}} C_{2}\bkt{\bf Adj} \sigma^2.
\ee
This leads to the  effective ccoupling
\be 
\frac{1}{g^2\bkt{\Lambda}}\ =\ \frac{1}{g_0^2}- \frac{5}{64 \pi^4 r^4}\bkt{\frac{ \bkt{\Lambda r}^2}{4}+\frac{3 \Lambda r}{2}+ 5 \log\bkt{\Lambda r}} C_{2}\bkt{\bf Adj}.
\ee

Here we see that in the decompactification limit the dependence on the energy scale vanishes. 
A similar computation for $d=9$ yields
\be 
\log\bkt{\Zv\Zh}\ =\ \frac{\Lambda r}{10}\bkt{\frac{\bkt{\Lambda r}^2}{3}\ + \frac{7\Lambda r}{2}+\frac{151}{6}} C_2\bkt{{\bf Adj}} \sigma^2\,,
\ee
which leads to following expression for effective coupling:
\be 
\frac{1}{g^2\bkt{\Lambda}}\ =\ \frac{1}{g_0^2}-\frac{\Lambda}{40 \pi ^5 r^4} \bkt{\frac{\bkt{\Lambda r}^2}{3}\ + \frac{7\Lambda r}{2}+\frac{151}{6}} C_2\bkt{{\bf Adj}} .
\ee
This is independent of the UV scale $\Lambda$ in the decompactification limit. The same calculation can be repeated for $d=10$ and it can be shown that the one-loop determinants do not contribute any divergences to the gauge coupling in the decompactification limit.

\section{Summary and discussions}
\label{s-summary}
In this note we have presented some nontrivial evidence in favor of the analytically continued expressions for one-loop determinants of supersymmetric gauge theories on $S^d$. We have computed the one-loop contribution to the gauge coupling as predicted from these determinants and showed that it is consistent with the well known one-loop behavior of minimally supersymmetric gauge theories in diverse dimensions. Unfortunately, at this point it is not clear  what are the actual Lagrangians for these gauge theories when they are on the sphere. Attempts to construct these theories by starting from ten dimensional SYM have not been successful so far. 

 Another possible way to find supersymmetric theories on spheres is to consider off-shell supergravity coupled to vector multiplets and then take the rigid limit, as in \cite{Festuccia:2011ws,Dumitrescu:2012ha}.   An off-shell formalism is known for $\NN=1$ supergravity in six dimensions, \cite{Bergshoeff:1985mz,Sokatchev:1988aa,Coomans:2011ih}.  This possibility was analyzed partially in \cite{Samtleben:2012ua} and it was shown that the off-shell theory with R-symmetry gauging does not admit $S^6$  as a supersymmetric background. It would be interesting to complete this analysis by determining whether or not the off-shell theory without R-symmetry gauging admits $S^6$  as a supersymmetric background, although in this case the supersymmetry might be enhanced to $\NN=2$.  For eight and nine dimensions we are unaware of an off-shell supergravity formulation.  
 
 It is reasonable to believe that the theories, assuming they exist, will have Lagrangians that have explicit terms that break the bosonic symmetries down to $SO(6)\simeq SU(4)$ in the six dimensional case, leaving an invariance under the $SU(4|1)$ superalgebra.  Likewise, we expect the eight and nine dimensional theories to only have an $SO(8)$ symmetry and be invariant under an $OSp(8|2)$ superalgebra. From the supergravity perspective, it would correspond to turning on non trivial background fields in addition to metric.

\section*{Acknowledgements}
 We thank Guido Festuccia and Luigi Tizzano  for discussions.   J.A.M also thanks the CTP at MIT  for 
hospitality during the course of this work.
The research of J.A.M.  is supported in part by
Vetenskapsr{\aa}det under grants \#2012-03269 and \#2016-03503 and by the Knut and Alice Wallenberg Foundation under grant Dnr KAW 2015.0083.
Work of U.N is supported by the U.S. Department of Energy under grant Contract Number de-sc0012567.

\bibliographystyle{JHEP}
\bibliography{refs}  
 
\end{document}